\documentclass[preprintnumbers,amsmath,amssymb]{revtex4}
\usepackage{dcolumn}
\usepackage{bm}
\usepackage[dvips]{color}
\usepackage[pdftex]{graphicx}

\newcommand{\beq}{\begin{equation}}
\newcommand{\eeq}{\end{equation}}
\newcommand{\bq}{\begin{equation}}
\newcommand{\eq}{\end{equation}}
\newcommand{\be}{\begin{equation}}
\newcommand{\ee}{\end{equation}}
\newcommand{\ba}{\begin{array}}
\newcommand{\ea}{\end{array}}
\newcommand{\beqa}{\begin{eqnarray}}
\newcommand{\eeqa}{\end{eqnarray}}
\newcommand{\bea}{\begin{eqnarray}}
\newcommand{\eea}{\end{eqnarray}}

\def\ol{\overline}

\def\[{\left[}
\def\]{\right]}
\def\({\left(}
\def\){\right)}

\evensidemargin=\oddsidemargin

\def\U{\mathcal{U}}
\def\OBZ{\mathcal{O}_{BZ}}
\def\OU{\mathcal{O}_\mathcal{U}}
\def\BZ{\mathcal{BZ}}
\def\OSM{\mathcal{O}_{\rm SM}}
\def\g5{\gamma_5}

\def\pslash{\not{\hbox{\kern-4pt $p$}}}
\def\qslash{\not{\hbox{\kern-4pt $q$}}}
\def\lv{\not{\hbox{\kern-4pt $L$}}}
\def\lsim{\mathrel{\raise.3ex\hbox{$<$\kern-.75em\lower1ex\hbox{$\sim$}}}}
\def\gsim{\mathrel{\raise.3ex\hbox{$>$\kern-.75em\lower1ex\hbox{$\sim$}}}}
\def\ifmath#1{\relax\ifmmode #1\else $#1$\fi}
\begin{document}

 \title{\large 
The large CP phase in  
$B_{s}-\ol{B}_{s}$ mixing
from primary scalar unparticles}

 \author{J.\ K.\ Parry}

 \affiliation{
Kavli Institute for Theoretical Physics China,\\
Institute of Theoretical Physics,\\
Chinese Academy of Science, Beijing 100190, China
\\
and \\
       Center for High Energy Physics,\\
       Tsinghua University, Beijing 100084, China
}
 \begin{abstract}
In this letter we consider 
the case of primary scalar unparticle contributions to $B_{d,s}$ mixing.
With particular emphasis on the impact of the recent hint of 
new physics in the measurement of the $B_s$ mixing phase, $\phi_s$, we
determine the allowed parameter space and impose bounds on 
the unparticle couplings.
 \\[2mm]
  PACS:  13.20.He  13.25.Hw  12.60.-i \hfill
 \end{abstract}
 \maketitle

\section{Introduction}

It has recently been suggested \cite{hep-ph/0703260,arXiv:0704.2457} that there
may exist a non-trivial scale invariant sector at high energies, 
known as unparticle stuff. These new fields with an 
infrared fixed point are called Banks-Zaks fields \cite{Nucl.Phys.B196.189},
interacting with Standard model fields via heavy particle exchange,
\bea
\frac{1}{M_\U^k}\OSM\OBZ.
\eea
Here $\OSM$ is a standard model(SM) operator of mass dimension $d_{\rm SM}$,
$\OBZ$ is a Banks-Zaks($\mathcal{BZ}$) operator of mass dimension 
$d_{\mathcal{BZ}}$, with $k=d_{\rm SM}+d_{\mathcal{BZ}}-4$,
 and $M_\U$ is the mass of the heavy particles mediating
the interaction. At a scale denoted by $\Lambda_\U$ the $\BZ$ operators
match onto unparticle operators with a new set of interactions,
\bea
C_\U \frac{\Lambda_\U^{d_\BZ-d_\U}}{M_\U^k}\OSM\OU
\eea
where $\OU$ is an unparticle operator with scaling dimension $d_\U$ 
and $C_\U$ is the coefficient of the low energy theory. Unparticle stuff
of scaling dimension $d_\U$ looks like a non-integral number $d_\U$ of
invisible massless particles. 
It was recently suggested in \cite{Grinstein:2008qk} that 
conformal invariance implies 
constraints on the scaling
dimension, $1\leq d_\U \leq 2$, 
and $3\leq d_\U \leq 4$, for scalar and vector unparticles respectively.
This also implies the constraint $2\leq d_\U \leq 3$ for the non-primary
operator $\partial^{\mu}\mathcal{O}_\U$.
We shall see later that assuming such 
constraints leads to the  contributions 
from vector unparticles and from the non-primary operator 
$\partial^{\mu}\mathcal{O}_\U$ being negligible, except in a small
region of parameter space where the scaling dimension 
approaches integer values.
As a result, in this work we shall follow the above 
suggestion and focus on the 
case of the primary 
scalar unparticle ($\OU$) 
with couplings to the SM quarks as follows,
\bea
{\mathcal{L}}_{S'}&=&
\frac{c_{L}^{S',\,q'q}}{\Lambda_\U^{d_\U-1}}\,
\ol{q}'(1-\g5)q\,\OU
+\frac{c_{R}^{S',\,q'q}}{\Lambda_\U^{d_\U-1}}\,
\ol{q}'(1+\g5)q\,\OU\label{S'_int}
\label{V_int}
\eea 
Here we assume that the left-handed and right-handed 
flavour-dependent dimensionless
couplings $c^{S',bq}_{L}$, $c^{S',bq}_{R}$, 
are independent parameters. The above notation, ${\mathcal{L}}_{S',bq}$,
has been used to distinguish the primary scalar operator studied here
from ${\mathcal{L}}_{S}$ for the non-primary scalar case 
$\partial^{\mu}\mathcal{O}_\U$ studied previously \cite{Parry:2008sr}.
We analyze a number of scenarios, in each case determining the 
allowed parameter space and placing bounds on the unparticle couplings.

The propagators for scalar 
unparticle are as
follows \cite{hep-ph/0703260,arXiv:0704.2588},
\bea
\int\,d^4 x \, e^{iP.x}
\langle 0| T\OU(x)\OU(0)|0\rangle&=&
i\frac{A_{d_\U}}{2\sin d_\U\pi}
\frac{1}{(P^2+i\epsilon)^{2-d_U}}e^{-i\phi_\U}\\
\eea
where
\bea
A_{d_\U}=\frac{16\pi^{5/2}}{(2\pi)^{2d_\U}}
\frac{\Gamma(d_\U+1/2)}{\Gamma(d_\U-1)\Gamma(2d_\U)},
\hspace{0.5cm}
\phi_\U=(d_\U-2)\pi
\eea

The effects of unparticles, both scalar and vector, on 
meson-anti-meson mixing has been studied in the literature 
\cite{arXiv:0705.1821,arXiv:0704.3532,arXiv:0705.0689,arXiv:0707.1234,arXiv:0707.1535,arXiv:0710.3663,arXiv:0711.3516,arXiv:0801.0895,Parry:2008sr,Zwicky:2007vv}.
In this work we shall study the constraints on 
primary scalar unparticles
coming from the measurements
of $B_{s,d}$ meson mass differences $\Delta M_{s,d}$ and also their 
CP violating phases $\phi_{s,d}$. In the $B_d$ system these quantities have
been well measured for some time and show no sizeable deviations from the 
SM expectations. In the $B_s$ system recent measurements have also found
small discrepancies between the SM expectation for $\Delta M_s$ 
\cite{hep-ex/0609040}, but now
the CP violating phase $\phi_s$, measured by the 
${\rm D}\emptyset$\cite{arXiv:0802.2255} 
and CDF\cite{arXiv:0712.2397} collaborations
reveals a deviation of 3$\sigma$\cite{arXiv:0803.0659,Lenz:2006hd}
\footnote{It should be noted that the combination of D$\emptyset$ and 
CDF results for $\phi_s$ were made 
without full knowledge of the likelihoods, which 
could have an effect on the significance of this SM deviation.
These are now being made available and a new experimental combination
is due to be released soon. Until that time this present hint of new physics 
is the best available and remains a very exciting prospect. }. 
This is the first evidence for new physics
in $b\leftrightarrow s$ transitions. 
Studying primary scalar unparticles, we 
derive the constraints imposed by these latest measurements on the 
coupling between SM fields and unparticles, with particular interest
on the impact of $\phi_s$.

\section{Meson-antimeson mixing from unparticles}

Using the interactions listed in eq.~(\ref{S'_int})
to evaluate the s- and t-channel contributions to meson mixing 
we obtain the effective Hamiltonian,
\bea
\mathcal{H}^{S',q'q}_{\rm eff}&=&
\frac{A_{d_\U}}{2\sin d_\U\pi}\frac{e^{-i\phi_{\U}}}{\Lambda_\U^{2d_\U-2}}
\left(\frac{1}{t^{2-d_\U}}+\frac{1}{s^{2-d_\U}}\right)
\left[
Q_{2}
\left(
c_L^{S',q'q}
\right)^2
+\tilde{Q}_{2}
\left(
c_R^{S',q'q}
\right)^2
+2\,Q_{4}
\left(
c_L^{S',q'q}c_R^{S',q'q}
\right)
\right]\label{Hscal'}
\eea
Here we have defined the quark operators $Q_1-Q_5$ and their 
hadronic matrix elements as in \cite{Parry:2008sr}.
Writing the $\Delta F=2$ 
effective Hamiltonian in terms of these operators we have,
\bea
{\mathcal H}_{\rm eff}^{q'q}
=\sum_{i=1}^5 
C^{S'}_i\,Q_i
+\sum_{j=1}^3 
\tilde{C}^{S'}_j\, \tilde{Q}_j
\eea
Here the operators 
$\tilde{Q}_{1,2,3}$ are obtained from $Q_{1,2,3}$ by the exchange
$L\leftrightarrow R$.
%
%
From eq.~(\ref{Hscal'}), it is straightforward to calculate
the Wilson coefficients for primary scalar unparticles which are as follows,
\bea
C^{S'}_2&=& \frac{A_{d_\U}}{\sin d_\U\pi}\frac{e^{-i\phi_{\U}}}{M_M^2} 
\left(\frac{M_M^2}{\Lambda_\U^2}\right)^{d_\U-1}
\left(
c_L^{S',q'q}
\right)^2
\\
C^{S'}_4&=& 2\frac{A_{d_\U}}{\sin d_\U\pi}\frac{e^{-i\phi_{\U}}}{M_M^2} 
\left(\frac{M_M^2}{\Lambda_\U^2}\right)^{d_\U-1}
\left(
c_L^{S',q'q}\,c_R^{S',q'q}
\right)
\\
\tilde{C}^{S'}_2&=& \frac{A_{d_\U}}{\sin d_\U\pi}\frac{e^{-i\phi_{\U}}}{M_M^2} 
\left(\frac{M_M^2}{\Lambda_\U^2}\right)^{d_\U-1}
\left(
c_R^{S',q'q}
\right)^2
\\
C^{S'}_1&=&C^{S'}_3=C^{S'}_5=\tilde{C}^{S'}_1=\tilde{C}^{S'}_3=0
\eea
where we have approximated $t=s\sim M_M^2$. This set of Wilson 
coefficients is rather similar to that for the non-primary
operator $\partial^{\mu}\mathcal{O}_\U$, with the main differences
being the sign of $C_2$ and the power to which the unparticle 
scale $\Lambda_\U$ is raised.
%
These Wilson coefficients will mix with each other as a result of 
renormalisation group(RG) running down to the scale of $M_M$.
For the B system, with a scale of new physics $\Lambda_\U=1$ TeV, 
these Wilson
coefficients at the scale $\mu_b=m_b$ are approximated as in 
\cite{Parry:2008sr}. 
%
The $\Delta F=2$ transitions are defined as,
\bea
\langle \ol{M}^0 |\mathcal{H}^{\Delta F=2}_{\rm eff}|M^0 \rangle
=M_{12}
\eea
with the meson mass eigenstate difference defined as,
$\Delta M \equiv M_H - M_L = 2|M_{12}|$.
We can define in a model independent way the contribution to meson
mixings in the presence of New Physics(NP) as,
\bea
M_{12}=M_{12}^{\rm SM}(1+R)
\label{totalM}
\eea
where $M_{12}^{\rm SM}$ denotes the SM contribution and 
$R\equiv r\,e^{i\,\sigma}=M_{12}^{\rm NP}/M_{12}^{\rm SM}$ 
parameterizes the NP contribution.
The associated CP phase may then be defined as,
\bea
\phi\equiv arg(M_{12})=\phi^{\rm SM}+\phi^{\rm NP}
\eea 
where $\phi^{\rm SM}=arg(M_{12}^{\rm SM})$ 
and $\phi^{\rm NP}=arg(1+r\,e^{i\,\sigma})$.

\subsection{$B_{s,d}$ mixing and unparticle physics}

In this work we shall focus on the constraints imposed on 
unparticle physics couplings from $B_{s,d}$ mixing. Therefore
we set $q'=b$, $q=s,\,d$ and $M^0=B^0_s,\,B^0_d$.
In the B system, the Standard Model contribution to $M_{12}^q$ is given by,
\be
M_{12}^{q,{\rm SM}}=\frac{G_F^2 M_W^2}{12 \pi^2}
M_{B_q}\hat{\eta}^B\,f_{B_q}^2 \hat{B}_{B_q}
(V_{tq}^* V_{tb})^2\, S_0(x_t)
\ee
where $G_F$ is Fermi's constant, $M_W$ the mass of the W boson,
$\hat{\eta}^B=0.551$ \cite{Buras:1990fn} 
is a short-distance QCD correction identical 
for both the $B_s$ and $B_d$ systems. The bag parameter $\hat{B}_{B_q}$ 
and decay constant $f_{B_q}$ are non-perturbative quantities and contain
the majority of the theoretical uncertainty. $V_{tq}$ and $V_{tb}$ are elements
of the Cabibbo-Kobayashi-Maskawa (CKM) matrix 
\cite{Cabibbo:1963yz,Kobayashi:1973fv}, 
and $S_0(x_t\equiv \bar{m}_t^2/M_W^2)=2.34\pm 0.03$, 
with $\bar{m}_t(m_t)=164.5\pm 1.1$ GeV \cite{:2008vn}, is one of the 
Inami-Lim functions \cite{Inami:1980fz}.

We can now constrain both the magnitude and phase of the NP
contribution, $r_q$ and $\sigma_q$, through the comparison of 
the experimental measurements with SM expectations. From the 
definition of eq.~(\ref{totalM}) we have the constraint,
\bea
\rho_q\equiv \frac{\Delta M_q}{\Delta M_q^{\rm SM}}
=\sqrt{1+2r_q\cos\sigma_q + r_q^2}
\label{rho}
\eea
The values for $\rho_q$ given by the UTfit analysis 
\cite{arXiv:0803.0659,arXiv:0707.0636} at the $95\%$ C.L. are,
\bea
\rho_d=\left[0.53,\,2.05\right],\hspace{1cm}
\rho_s=\left[0.62,\,1.93\right]
\eea
These constraints on $\rho_q$ encode the CP conserving measurements
of $\Delta M_{d,s}$.
The phase associated with NP can also be 
written in terms of $r_q$ and $\sigma_q$,
\bea
\sin\phi_q^{\rm NP}=\frac{r_q\sin\sigma_q}
{\sqrt{1+2 r_q\cos\sigma_q+r_q^2}}
\label{phiNP}
\eea
Here \cite{arXiv:0803.0659,arXiv:0707.0636} gives the $95\%$ C.L. constraints,
\bea
\phi_d^{\rm NP}&\hspace{-1mm}=&\hspace{-1mm}
\left[-16.6,\,3.2\right]^{\rm o}\label{phidNP}\\
\phi_{s}^{\rm NP}&\hspace{-1mm}=&\hspace{-1mm}
\left[-156.90,\,-106.40\right]^{\rm o}\cup
\left[-60.9,\,-18.58\right]^{\rm o}
\label{phisNP}
\eea
these constraint represent those of the CP phase measurements of $\phi_{d,s}$.
As in \cite{Parry:2008sr}, 
in order to consistently apply the above constraints all input 
parameters are chosen to match those used in the analysis
of the UTfit group \cite{arXiv:0803.0659,arXiv:0707.0636}.

\section{Numerical Analysis}

\begin{figure}[h]
\includegraphics[width=5.5truecm,clip=true]{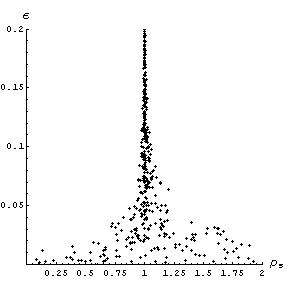}
\includegraphics[width=5.5truecm,clip=true]{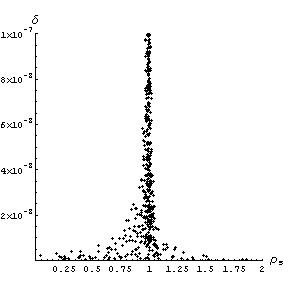}
\caption{Size of $\rho_{s}$ against $\epsilon$(left) and 
$\delta$(right) for unparticle operator
$\partial^{\mu}\mathcal{O}_\U$
with real couplings $c_L^{S',bs},\,c_R^{S',bs}\in (-1,1)$ and 
unparticle
scale $\Lambda_\U=1$ TeV.}
\label{fig1}
\end{figure}

\begin{figure}[h]
\includegraphics[width=5.5truecm,clip=true]{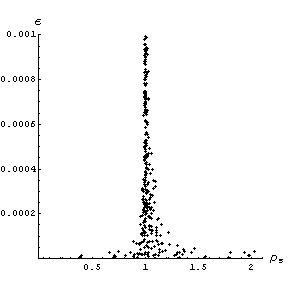}
\includegraphics[width=5.5truecm,clip=true]{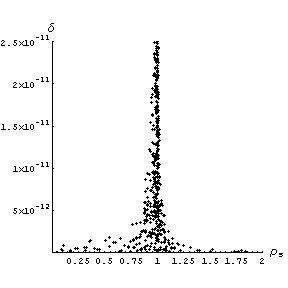}
\caption{Size of $\rho_{s}$ against $\epsilon$(left) and 
$\delta$(right) for unparticle operator
$\mathcal{O}^{\mu}_\U$
with real couplings $c_L^{V,bs},\,c_R^{V,bs}\in (-1,1)$ and
unparticle
scale $\Lambda_\U=1$ TeV.}
\label{fig2}
\end{figure}

First let us re-examine the contributions to 
$B_{d,s}$ mixing from primary vector unparticles $\mathcal{O}^{\mu}_\U$ and
the non-primary $\partial^{\mu}\mathcal{O}_\U$. The Wilson coefficients
for these cases can be found for example 
in \cite{Parry:2008sr}. Following the 
arguments of \cite{Grinstein:2008qk} we have the bounds 
$2\leq d_\U \leq 3$ for $\partial^{\mu}\mathcal{O}_\U$ and 
$3\leq d_\U \leq 4$ for $\mathcal{O}^{\mu}_\U$.
These bounds produce a very large suppression of the contributions
to $B_{d,s}$ mixing from these unparticle operators, unless we are 
sufficiently close to 
the pole in the unparticle propagator. These poles occur 
as $d_\U$ approaches integer values, due the $1/\sin  d_\U\pi$ 
nature of the propagator. Therefore the only significant 
contributions may occur for values of the scaling dimension,
$d_\U=A+\epsilon$ and $d_\U=B-\delta$, where $A=2(3)$ and $B=3(4)$
for $\partial^{\mu}\mathcal{O}_\U$($\mathcal{O}^{\mu}_\U$), with 
$\epsilon,\delta\ll 1$. Fig.~\ref{fig1} and \ref{fig2} show the 
size of $\rho_s$ which results from varying $\epsilon$ and $\delta$
for couplings between SM operator and unparticle operator in the 
range $(-1,1)$. From these plots we can see that for the unparticle 
operator $\partial^{\mu}\mathcal{O}_\U$, we require 
$\epsilon \lesssim 0.1$ or $\delta \lesssim 10^{-7}$ to get significant 
contributions to $B_s$ mixing. For the vector unparticle operator
$\mathcal{O}^{\mu}_\U$ the requirement is 
$\epsilon \lesssim 0.0005$ or $\delta \lesssim 10^{-10}$. 
The range of values of $d_\U$ for which the present allowed 
region for $\rho_s$ may provide a constraint on the couplings
is even smaller than those stated above. For example for 
the non-primary operator $\partial^{\mu}\mathcal{O}_\U$ we require
$\epsilon\lesssim 0.025$ before the $\rho_s$ constraints begin 
to have any effect on the size of the couplings.
The requirements 
for non-negligible contributions to $B_d$ mixing will be less 
strict. 
It seems that, baring a small region of parameter space, the 
contributions from primary vector and non-primary scalar unparticles
are negligible. As a result we here consider only the primary 
scalar unparticle contribution to $B_{s,d}$ mixing.

For our analysis of the primary scalar contribution to $B_{s,d}$
mixing we shall take the unparticle
scale $\Lambda_\U=1$ TeV and generally fix $d_\U=\frac{3}{2}$. 
In the following we shall consider 
possible coupling patterns as follows,
\begin{itemize}
\item { One real coupling}; $c^{S',bq}_L\neq 0$ and $c^{S',bq}_R=0$, with $c^{S',bq}_L \in \mathbb{R}$
\item { Two real couplings}; $c^{S',bq}_L\neq 0$ and $c^{S',bq}_R\neq 0$, with $\{c^{S',bq}_L,c^{S',bq}_R\}\in\mathbb{R}$
\item { One complex coupling}; $c^{S',bq}_L\neq 0$ and $c^{S',bq}_R=0$, with $c^{S',bq}_L\in\mathbb{C}$
\item { Two complex couplings}; $c^{S',bq}_L\neq 0$ and $c^{S',bq}_R\neq 0$, with $\{c^{S',bq}_L,c^{S',bq}_R\}\in\mathbb{C}$
\end{itemize}

\begin{figure}[h]
\includegraphics[width=6truecm,clip=true]{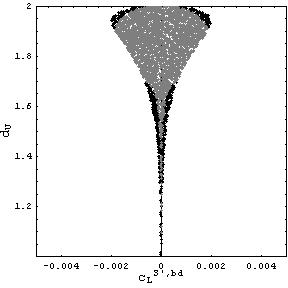}
\includegraphics[width=6truecm,clip=true]{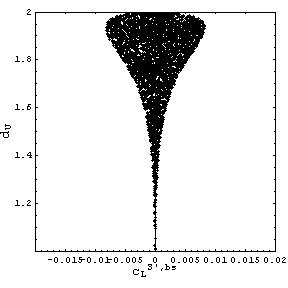}
\caption{Constraints on the $d_\U$ versus $c_L^{S',bq}$ parameter space 
from $B_d$ mixing(left) and $B_s$ mixing(right) for the case of 
a single real coupling $c_L^{S',bq}\neq 0$ and $c_R^{S',bq}=0$. 
Black points indicate the 
$\Delta M_{d,s}$ allowed regions, while grey points indicate the regions
are in agreement with both $\Delta M_{q}$ and the CP phase $\phi_q$.}
\label{fig3}
\end{figure}

\begin{figure}[h]
\includegraphics[width=6truecm,clip=true]{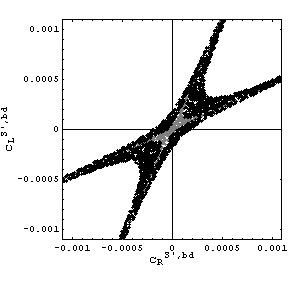}
\includegraphics[width=6truecm,clip=true]{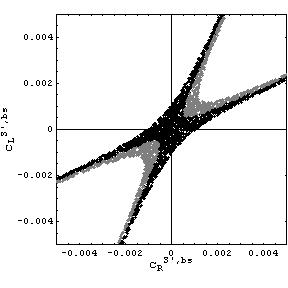}
\caption{Plot of the allowed $c_L^{S',bq}$-$c_R^{S',bq}$ parameter space 
 for $B_d$ mixing(left) and $B_s$ mixing(right) 
in the case of two real couplings and scaling dimension fixed as 
$d_\U=3/2$. 
Black points show regions
which agree with the measurement of $\Delta M_{d,s}$ while grey points show
additional agreement with the measurement of the CP phases $\phi_{d,s}$.}
\label{fig4}
\end{figure}

In the first case, a single real coupling 
($c^{S',bq}_L\neq 0$, $c^{S',bq}_R=0$) between primary scalar
unparticle operator and our SM quark operator, we allow the 
scaling dimension to vary in the range indicated by the bounds discussed
earlier. Fig.~\ref{fig3} and \ref{fig4} show the experimentally 
allowed parameter space in this case in the plane of the scaling 
dimension, $d_\U$, and the coupling $c_L^{S',bq}$. This plot shows a very
similar behaviour to that found for non-primary scalar
unparticles with real couplings $c_L^{S,bq}=c_R^{S,bq}$ \cite{Parry:2008sr},
with the allowed parameter space
generally increasing for increasing $d_\U$. This similarity between these
two seemingly different cases is simply because they both give 
negative $M_{12}^{\U}$ due to the sign of the Wilson coefficient $C_2$.

Fig.~\ref{fig3} shows the allowed $d_\U-c_L^{S',bq}$ parameter space, 
black points obey the constraint from $\Delta M_d$ while grey
points obey constraints from both $\Delta M_d$ and $\phi_d$.
From the left panel of fig.~\ref{fig3}, with $d_\U=\frac{3}{2}$,
 we can extract bounds,
\bea
&|c_L^{S',bd}|&\leq 0.0002\hspace{1.15cm}(\Delta M_d\,{\rm only})\\
&|c_L^{S',bd}|&\leq 0.00005\hspace{0.975cm}(\Delta M_d\,\&\, \phi_d)
\eea
From the right panel of fig.~\ref{fig3}, with $d_\U=\frac{3}{2}$, 
we can extract the bounds,
\bea
&|c_L^{S',bs}|&\leq 0.001\hspace{1.15cm}(\Delta M_s\,{\rm only})\\
&|c_L^{S',bs}|&\,\,excluded\hspace{0.85cm}(\Delta M_s\,\&\, \phi_s)
\eea
where we see that the combined constraints of $\Delta M_s$ \& $\phi_s$
exclude this possibility at the $3\,\sigma$ level, for all values of 
$d_\U$.

\begin{figure}[h]
\includegraphics[width=6truecm,clip=true]{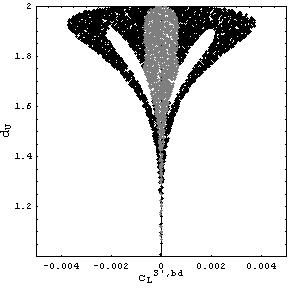}
\includegraphics[width=6truecm,clip=true]{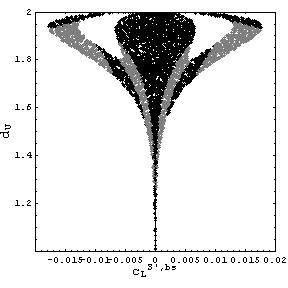}
\caption{Variation of allowed parameter space 
of the real coupling $c_L^{S',bq}=c_R^{S',bq}$ 
with scaling dimension $d_\U$ for $B_d$ mixing (left) and $B_s$ mixing (right).
Black plotted points agree with the CP conserving mixing quantities
$\Delta M_{d,s}$, while grey points also agree with the CP phases 
$\phi_{d,s}$.}
\label{fig5}
\end{figure}

For the second case of two real couplings, $c^{S',bq}_L$ and $c^{S',bq}_R$, we
take two sub-cases, $c^{S',bq}_L\neq c^{S',bq}_R$ 
and $c^{S',bq}_L=c^{S',bq}_R$,
shown in fig.~\ref{fig4} and \ref{fig5} respectively.
For the first sub-case, $c^{S',bq}_L\neq c^{S',bq}_R$, 
shown in fig.~\ref{fig4},
there are generally 
no bounds that can be set on the couplings with the allowed parameter space
stretching along two lines, similar to the case of two real non-equal 
couplings to unparticle operator $\partial^{\mu}\mathcal{O}_\U$. 
Looking at the grey area of fig.~\ref{fig4}
we can see that including the $\phi_s$ and $\phi_d$ constraints 
further restrict the allowed parameter space. 
For the second sub-case, $c^{S',bq}_L=c^{S',bq}_R$, shown in fig.~\ref{fig5}, 
the allowed parameter space is rather similar to the case of
unparticle operators $\mathcal{O}_\U^{\mu}$ and $\partial^{\mu}\mathcal{O}_\U$ 
with a single real coupling, \cite{Parry:2008sr}.
For this sub-case we can now set the bounds, 
\bea
&|c_L^{S',bd}|=|c_R^{S',bd}|&\leq 0.00034\hspace{1cm}(\Delta M_d\,{\rm only})\\
&|c_L^{S',bd}|=|c_R^{S',bd}|&\leq 0.00012\hspace{1cm}(\Delta M_d\,\&\, \phi_d)\\
&|c_L^{S',bs}|=|c_R^{S',bs}|&\leq 0.0013\hspace{1.15cm}(\Delta M_s\,{\rm only})\\
0.00058 \leq &|c_R^{S',bs}|=|c_L^{S',bs}|&\leq 0.0013\hspace{1.15cm}(\Delta M_s\,\&\, \phi_s)
\eea
for $d_\U=\frac{3}{2}$.

\begin{figure}[h]
\includegraphics[width=6truecm,clip=true]{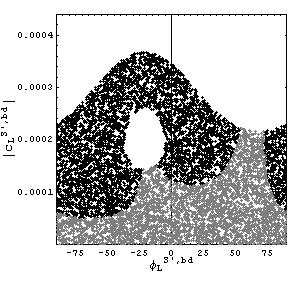}
\includegraphics[width=6truecm,clip=true]{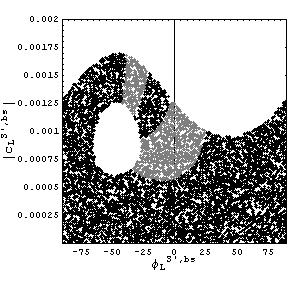}
\caption{Plot of the allowed $|c_L^{S',bq}|$-$\phi_L^{S',bq}$ parameter space 
 for $B_d$ mixing (left) and $B_s$ mixing (right) 
for the case of a single complex coupling
$c_L^{S',bq}\neq 0$ and $c_R^{S',bq}=0$ and scaling dimension fixed as 
$d_\U=3/2$.
Black plotted points agree with the CP conserving mixing quantities
$\Delta M_{d,s}$, while grey points also agree with $\phi_{d,s}$.}
\label{fig6}
\end{figure}

The allowed parameter space for the third case, 
a single complex coupling, 
$c^{S',bq}_L=|c_L^{S',bq}|\,{\rm e}^{i\phi_L^{S',bq}}$
 and $c^{S',bq}_R=0$, is shown in fig.~\ref{fig6}. In this case we have 
the bounds, 
\bea
&|c_L^{S',bd}|\leq 0.00029&\hspace{1.15cm}(\Delta M_d\,{\rm only})\\
&|c_L^{S',bd}|\leq 0.00018&\hspace{1.15cm}(\Delta M_d\,\&\, \phi_d)\\
&|c_L^{S',bs}|\leq 0.0014&\hspace{1.15cm}(\Delta M_s\,{\rm only})\\
&0.00046\leq |c_L^{S',bs}|\leq 0.0014
\hspace{0.35cm}
{\rm and }
\hspace{0.35cm}
48.2^{o}\leq \phi^{S',bs}_L\leq 115.6^{o}&
\hspace{1.15cm}(\Delta M_s\,\&\, \phi_s)
\eea

\begin{figure}[h]
\includegraphics[width=6truecm,clip=true]{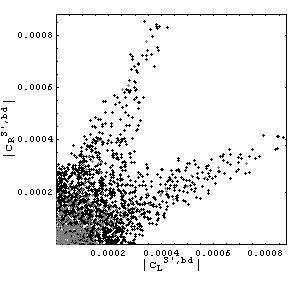}
\includegraphics[width=6truecm,clip=true]{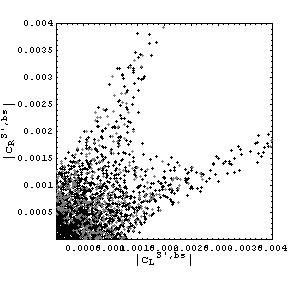}
\caption{Plot of the allowed $|c_L^{S',bq}|$-$|c_R^{S',bq}|$ parameter space 
 for $B_d$ mixing(left) and $B_s$ mixing(right) 
for the case of two complex couplings
$c_L^{S',bq}\neq 0$ and $c_R^{S',bq}\neq 0$.
Black plotted points agree with the CP conserving quantities
$\Delta M_{d,s}$, while grey points also agree with $\phi_{d,s}$.}
\label{fig7}
\end{figure}

\begin{figure}[h]
\includegraphics[width=6truecm,clip=true]{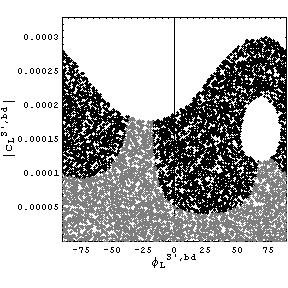}
\includegraphics[width=6truecm,clip=true]{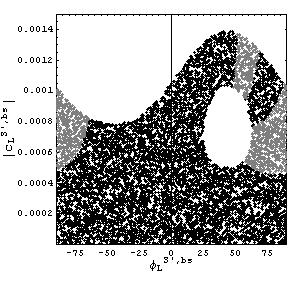}
\caption{Plot of the allowed $|c_L^{S',bq}|$-$\phi_L^{S',bq}$ parameter space 
 for $B_d$ mixing(left) and $B_s$ mixing(right) 
for the case of one complex coupling $c_L^{S',bq}=c_R^{S',bq}$
and scaling dimension fixed as 
$d_\U=3/2$.
Black plotted points agree with the CP conserving mixing quantities
$\Delta M_{d,s}$, while grey points also agree with $\phi_{d,s}$.}
\label{fig8}
\end{figure}

For the final case of two complex couplings, 
$c^{S',bq}_L=|c_L^{S',bq}|\,{\rm e}^{i\phi_L^{S',bq}}$
and $c^{S',bq}_R=|c_R^{S',bq}|\,{\rm e}^{i\phi_R^{S',bq}}$, we again 
consider two sub-cases, 
$c^{S',bq}_L\neq c^{S',bq}_L$
and 
$c^{S',bq}_L=c^{S',bq}_L$. The first sub-case, 
$c^{S',bq}_L\neq c^{S',bq}_L$, is shown in fig.~\ref{fig7} where the 
allowed parameter space is again stretched along two lines. This
is again similar to the parameter space found for the operators 
$\partial^{\mu}\mathcal{O}_\U$ and $\mathcal{O}^{\mu}_\U$.
From the left panel of fig.~\ref{fig7} we can see that again the 
inclusion of the CP constraint from $\phi_d$ decreases the allowed region
further. The right panel shows that CP phase $\phi_s$ has the same effect
and further disfavours a small region near the origin. Fig.~\ref{fig8}
displays the allowed parameter space for the sub-case 
$c^{S',bq}_L=c^{S',bq}_L$ where we can impose the bounds,
\bea
&|c_L^{S',bd}|=|c_R^{S',bd}|\leq 0.00037&\hspace{1.15cm}(\Delta M_d\,{\rm only})\\
&|c_L^{S',bd}|=|c_R^{S',bd}|\leq 0.00022&\hspace{1.15cm}(\Delta M_d\,\&\, \phi_d)\\
&|c_L^{S',bs}|=|c_R^{S',bs}|\leq 0.0017&\hspace{1.15cm}(\Delta M_s\,{\rm only})\\
&0.00056\leq |c_L^{S',bs}|=|c_R^{S',bs}|\leq 0.0017
\hspace{0.35cm}
{\rm and }
\hspace{0.35cm}
-41.8^{o}\leq \phi^{S',bs}_L\leq 25.6^{o}&
\hspace{1.15cm}(\Delta M_s\,\&\, \phi_s)
\eea
for a scaling dimension of 
$d_\U=3/2$

\section{Conclusion}

It was recently suggested in \cite{Grinstein:2008qk} that 
conformal invariance implies 
constraints on the scaling
dimension, $1\leq d_\U \leq 2$, 
and $3\leq d_\U \leq 4$, for scalar and vector unparticles respectively.
These constraints imply a rather large suppression of the contribution
of vector and non-primary scalar unparticles to $B_{d,s}$ mixing, 
except for in a rather small range of scaling dimension close to 
integer values. The largest range where non-negligible contributions
exist is for the non-primary operator $\partial^{\mu}\mathcal{O}_\U$
with $2\leq d_\U\lesssim 2.1$. For primary scalar unparticles the suppression
is far less and as such the range of scaling dimension for which sizable
contributions to $B_{d,s}$ mixing exists is far larger. As a result, we
have here considered, for the first time, the contribution of a
primary scalar unparticle operator $\mathcal{O}_\U$ to $B_{d,s}$ mixing.
Considering a number of different coupling patterns, we have determined the 
allowed parameter space and set bounds on the unparticle couplings in 
each case. To illustrate the impact of the inclusion of the constraints
from the CP phases associated with $B_{d,s}$ mixing, in particular the 
recently measured hint of new physics in $\phi_s$, we have 
analyzed constraints with and without the inclusion of $\phi_{d,s}$. 
In each case we 
have found that these CP phases have an important role to play in constraining
the parameter space, highlighted by the case 
of a single real unparticle coupling, $c_L^{S',bs}\neq 0$
and $c_R^{S',bs}= 0$, which was 
excluded for all values of the scaling dimension $d_\U$.

\end{document}